\documentclass[aps,pre,twocolumn,groupedaddress,showpacs,10pt,showkeys]{revtex4-1}
\usepackage{graphicx}
\usepackage{bm}
\usepackage{hyperref}
\usepackage[mathlines]{lineno}
\usepackage{amsmath}
\usepackage{mathtools}

\begin{document}


\title{Small-angle scattering from fat fractals}

\author{E. M. Anitas}
\email[]{anitas@theor.jinr.ru}
\affiliation{Joint Institute for Nuclear Research, Dubna 141980, Moscow region, Russian Federation}
\affiliation{Horia Hulubei
National Institute of Physics and Nuclear Engineering, RO-077125 Bucharest-Magurele, Romania}

\date{\today}

\begin{abstract}
A number of experimental small-angle scattering (SAS) data are characterized by a succession of power-law decays with arbitrarily decreasing values of scattering exponents. To describe such data, here we develop a new theoretical model based on $\mathrm3D$ fat fractals (sets with fractal structure, but nonzero volume) and show how one can extract structural information about the underlying fractal structure. We calculate analytically the monodisperse and polydisperse SAS intensity (fractal form factor and structure factor) of a newly introduced model of fat fractals and study its properties in momentum space. The system is a $3D$ deterministic mass fractal built on an extension of the well-known Cantor fractal. The model allows us to explain a succession of power-law decays and respectively, of generalized power-law decays (superposition of maxima and minima on a power-law decay) with arbitrarily decreasing scattering exponents in the range from zero to three. We show that within the model, the present analysis allows us to obtain the edges of \textit{all} the fractal regions in the momentum space, the number of fractal iteration and the fractal dimensions and scaling factors \textit{at each} structural level in the fractal. We applied our model to calculate an analytical expression for the radius of gyration of the fractal. The obtained quantities characterizing the fat fractal are correlated to variation of scaling factor with the iteration number.
\end{abstract}


\maketitle

\section{\label{sec:intro}{Introduction}}

The small-angle scattering (SAS; X-rays, neutrons, light)~\cite{glatter82:book,svergun87:book} has been established as a powerful experimental technique for structural investigations of various types of disordered systems (biological, polymeric) at nano- and microscale. The technique yields the differential elastic cross section per unit solid angle as a function of the momentum transfer which describes, through a Fourier transform, the spatial density-density correlations of the system. Since a large class of systems show the property of self-similarity  across the scales, the concept of fractal geometry~\cite{mandelbrot83:book, gouyet96:book} is very useful in modeling their structure and in describing the correlations between the microscopic and macroscopic properties. The effectiveness of the SAS method in investigating the fractal microstructure arise from the ability to differentiate between surface and mass fractals~\cite{martin87,schmidt91}. The difference is accounted through the value of the scattering exponent of the power-law decay of SAS intensity in the fractal region, with $I(q)~\propto~q^{-\tau}$, where $\tau = \mathrm{D}_{\mathrm{m}}$ for mass fractals and $\tau = 6-\mathrm{D}_{\mathrm{s}}$ for surface fractals. Here $\mathrm{D}_{\mathrm{m}}$ and $\mathrm{D}_{\mathrm{s}}$ are the mass and, respectively, surface fractal dimension~\cite{mandelbrot83:book} and lie within $0<\mathrm{D}_{\mathrm{m}}<3$ for a mass fractal~\cite{chernyJACR10}, and within $2<\mathrm{D}_{\mathrm{s}}<3$ for a surface fractal~\cite{martin87,schmidt91}.

Experimental SAS data can show a \textit{succession} of mass and/or surface fractal power-law regions, whose scattering exponents take arbitrarily decreasing values~\cite{zhao09,headen09,golosova12} and the existing theoretical models either provide an insufficient microstructural description from this type of SAS data, either they can not describe the full spectrum (note that the case of increasing values of scattering exponents can be explained in the framework of multi-phase systems~\cite{chernyArxiv}). SAS data modeled by the classical Beaucage model~\cite{beaucage95} can explain only a succession of power-law decays and involve a generic hierarchical structure linked to the fractal power-law regime. It gives fractal dimensions and the size of each structural, together with the specific surface when the Porod region ($I(q)~\propto~q^{-4}$) is present. These parameters provide important informations about the spatial organization of the fractals, but a more complete characterization is needed since to a large number of structures may correspond a given set of fractal dimensions. In addition, recent technological progress has allowed the development of deterministic fractal structures at nano/micro scales~\cite{cnar08, newkome06, berenschot13}. These type of structures are characterized by a generalized power-law decay (superposition of maxima and minima on a power-law decay, with the scattering exponent equal to the fractal dimension of the fractal) in momentum space and therefore due to log-oscillations, additional information can be obtained about the fractals describing the hierarchical structures, such as fractal iteration number, scaling factor and the number of structural units of which the fractal is composed, thus greatly improving our understanding on their structural properties~\cite{chernyJACR10,chernyPRE11}. 

To explain such \textit{a succession} of (generalized) power-law decays and illustrate the SAS properties we have calculated analytically the fractal form and structure factor from a system of randomly oriented, non-interacting, monodisperse and polydisperse deterministic $3D$ fat fractals (in mathematics known also as $\epsilon$-Cantor sets~\cite{aliprantis98:book}), which are sets with fractal structure, but nonzero volume (positive Lebesgue measure). The method is based on the theoretical approach which was successfully employed to describe SAS from thin fractals (known in literature simply as fractals)~\cite{schmidt86,chernyJSI,chernyJACR10,chernyPRE11}. In the case of monodisperse fractals it gives a generalized power-law decay. The polydispersity smooths the scattering intensity and leads to the simple power-law behavior observed in experimental data. The fat fractal system suggested here is built by a set of iterative rules, with the scaling factor increased as a function of the iteration number which, in turn, give rise to various fractal regions with different lengths and scattering exponents.

In this paper, it is shown that scattering intensity from deterministic fat fractals includes successive fractal regions with arbitrarily decreasing values of the scattering exponents and allows us to take full advantage of the properties of deterministic thin fractals~\cite{chernyJACR10,chernyPRE11}. We derive analytically the main properties in momentum space: fractal form and structure factor, and explain how to extract the main structural characteristics of mono and polydisperse fat fractals from SAS data. In particular, we focus on determining the edges of the fractal regions, fractal dimensions and scaling factors at each structural level, and the fractal radius of gyration.

\section{\label{sec:theory}{Theoretical background}}

\subsection{Fat fractals}

Fat fractals are characterized by the dependence of their apparent size on the scale resolution and they are quite different from the familiar thin fractals. To make the distinction between fat and thin fractals more clear, we consider the well-known $3D$ Cantor set~\cite{chernyJACR10}. In the later case, the initial cube ($m=0$; m being the fractal iteration number) is divided into 27 parts, the eight cubes are left in the corners ($m=1$), with side length $1/3$ from the initial cube, and the 19 parallelepipeds are removed. Then we repeat the same operation on each of the remaining eight cubes, thus leaving 64 cubes of side length $1/3^{2}$ ($m=2$) and so on. The thin Cantor fractal is obtained in the limit $m \rightarrow \infty$ and has zero volume (Lebesgue measure) and fractal dimension $\log{8}/\log{3}$. The "fattened" version of this thin fractal is obtained by keeping the cubes instead of side length $1/3$ ($m=1$), then $1/3^{2}$ ($m=2$), then $1/3^{3}$ ($m=3$), etc. The resulting fractal is topologically equivalent to the thin Cantor fractal, but the holes decrease in size sufficiently fast so that, when $m \rightarrow \infty$, the fractal has nonzero and finite volume, and fractal dimension 3 (see below). The resolution dependent volume $V(\epsilon)$ can be calculated by covering the fractal with balls of size $\epsilon$. Then the volume can be written~\cite{farmer85}
\begin{equation}
V(\epsilon) \approx V(0)+A\epsilon^{\eta},
\label{eq:volumes}
\end{equation}
where $A$ is a constant which depends on the units used and $V(0)$ is the  volume in the limit $\epsilon \rightarrow 0$. Using Eq.~(\ref{eq:volumes}) one can define the scaling exponent $\eta$ in the following way~\cite{farmer85}
\begin{equation}
\eta=\lim_{\epsilon \rightarrow 0}\frac{\log(V(\epsilon)-V(0))}{\log(\epsilon)},
\label{eq:scalingexponent}
\end{equation}
where, by definition $0 \leq \eta \leq \infty$ ($\eta$ is equal to $\infty$ for non-fractal sets and is finite for fractal sets) and provides a useful way to quantifies the fractal properties as opposed to the fractal dimension, since the fat fractal definition implies they have an integer fractal dimension. Although it is an essential parameter from which we can distinguish fat fractals ($\eta$ is independent of $d$) from thin fractals ($\eta=3-d$)~\cite{farmer85,umberger85} where $d$ is the fractal dimension, the connection between this scaling exponent and the small-angle scattering is beyond the scope of this paper. 

\subsection{Small-angle scattering}

We revise in this Section the theoretical formalism of SAS scattering (neutron, X-ray, light, or electron diffraction) from a two-phase sample consisting of microscopic objects with the scattering length $b_j$ and scattering length density (SLD) $\rho_\mathrm{m}$ immersed into a solid matrix of SLD $\rho_\mathrm{p}$, and neglect multiple scattering. Then the total cross section is given by~\cite{svergun87:book} $\mathrm{d}\sigma/\mathrm{d}\Omega=|A(\bm{q})|^2$, where $A(\bm{q})\equiv \int_{V'} \rho_\mathrm{s}(\bm{r}) e^{i \bm{q}\cdot\bm{r}}\mathrm{d}^3 r$ is the total scattering amplitude and $V'$ is the total volume irradiated by the incident beam. The SLD can be defined with the help of Dirac's $\delta$-function as $\rho_\mathrm{s}(\bm{r})=\sum_j b_j\delta(\bm{r}-\bm{r}_j)$, where $\bm{r}_j$ are the microscopic object positions.

In practice, it is convenient to represent the total scattering amplitude as a sum of amplitudes of rigid objects. For instance, considering the scattering from stiff fractals, whose spatial positions and orientations are uncorrelated, one can choose them as the objects. Then the scattering intensity (that is, the cross section per unit volume of the sample) is given by 
\begin{equation}
I(q) = n |\Delta\rho|^{2} V^{2}\left\langle \left|F(\bm{q})\right|^{2}\right\rangle, 
\label{eq:intensitygeneral}
\end{equation}
where $n$ is the fractal concentration, $V$ is the volume of each fractal, $\Delta\rho = \rho_\mathrm{m}-\rho_\mathrm{p}$ is the scattering contrast and $F(\bm{q})$ is the normalized form factor
\begin{equation}
F(\bm{q})=\frac{1}{V}\int_{V}e^{-i\bm{q}\cdot\bm{r}}\mathrm{d}\bm{r},
\label{eq:formfactorgeneral}
\end{equation}
obeying the condition $F(0) \equiv 1$. The brackets $\left\langle \cdots \right\rangle$ stand for the ensemble averaging over all orientations of the fractals. If the probability of any orientation is the same, then it can be calculated 
by averaging over all directions $\bm{n}$ of the momentum transfer $\bm{q}=q \bm{n}$, 
that is, by integrating over the solid angle in the spherical coordinates ${q}_{x}=q 
\cos\varphi \sin\vartheta$, ${q}_{y}=q \sin\varphi \sin\vartheta$ and ${q}_{z}=q 
\cos\vartheta$ 
\begin{equation}
\langle f(q_x,q_y,q_z) \rangle\equiv\frac{1}{4\pi}\int_{0}^{\pi}\mathrm{d}\vartheta\sin\vartheta\int_{0}^{2\pi}\mathrm{d}
\varphi\,f(q,\vartheta,\varphi).
\label{aver}
\end{equation}

Once a deterministic fractal is composed of $N_m$ objects (e.g. of the same radius $R$), then the form factor can be written as
\begin{equation}
F(\bm{q})=\rho_{\bm{q}}F_{0}(\bm{q}R)/N_{m},
\label{eq:fractalff}
\end{equation}
where $\rho_{\bm{q}}=\sum_{j}e^{-i \bm{q}\bm{r}_{j}}$ is the Fourier transform of the density of ball centers, $\bm{r}_{j}$ are the center-of-mass positions of balls and $m$ is the iteration number. Then, by using Eq.~(\ref{eq:intensitygeneral}), the scattering intensity becomes~\cite{chernyPRE11}
\begin{equation}
I(q)=I(0)S(q)|F_{0}(\bm{q}R)|^2/N_{m} \label{eq:intsq},
\end{equation}
where $I(0)= n |\Delta\rho|^{2} V^{2}$ is the intensity in zero angle, $F_{0}({\bm{q}}R)$ is the subunit form factor and $S(q)$ is the fractal structure factor defined by
\begin{equation}
S(q)\equiv \langle \rho_{\bm{q}}\rho_{-\bm{q}} \rangle /N_{m}.
\label{eq:sfdef}
\end{equation}
The choice of the subunit form factor $F_{0}(\bm{q}R)$ and of the fractal structure factor $S(q)$ is rather arbitrarily and depends on the shape of the scattering units and, respectively, on their relative positioning.

In a physical system scatterers almost always have different sizes. Therefore, a more realistic description should involve size polydispersity. Here we consider an ensemble of fractals with various sizes and forms. The distribution function $D_\mathrm{N}(l)$ of the scatterer sizes is defined in such a way that $D_\mathrm{N}(l)dl$ gives the probability of finding a fractal whose size falls within the interval $(l,l+dl)$. Specifically, we choose the log-normal distribution (see Ref.~\cite{chernyPRE11} for details) where the mean length $l_0$ and its relative variance  $\sigma_\mathrm{r}$ are given by
\begin{equation}
l_{0}\equiv\left\langle l\right\rangle_{D},\quad \sigma_\mathrm{r}\equiv\big(\langle 
l^{2}\rangle_{D}-l_{0}^{2}\big)^{1/2}/l_{0}, \label{eq:lengthandvariance}
\end{equation}
Thus, the average in Eq.~(\ref{eq:intensitygeneral}) is taken both over angles and sizes. Polydispersity obviously smears the intensity curves, and the oscillations become smoother~\cite{chernyPRE11}. 

\section{\label{construction}{Construction of the fat fractal}}

The scattering exponents $\tau$ in SAS intensities, as already mentioned before, are related to the fractal dimensions of the system. Therefore, in order to describe successive power-law regimes with decreasing values of the scattering exponents, the model based on deterministic fractals shall be specified by taking into consideration increasing values of the scaling factors with the iteration number, not after each iteration, but \textit{after a given number of iterations} (every second, every third iteration etc).

The construction process of the fat fractal, embedded into $3D$ space, is very similar to that of mass generalized thin Cantor fractals~\cite{chernyJACR10} and of mass generalized thin Vicsek fractals~\cite{chernyPRE11}. One follow a top-down approach in which an initial structure is repeatedly divided into a set of smaller structures of the same type, according to a given rule~\cite{hamburgerlidar96}. One and the same rule is kept from one iteration to another but the scaling factor is increased after every second iteration.

We start with a cube of edge $l_0$ (called zero-order iteration or initiator) and specify it in Cartesian coordinates as a set of points satisfying the conditions $-l_{0}/2 \leq x \leq l_{0}/2$, $-l_{0}/2 \leq y \leq l_{0}/2$, $-l_{0}/2 \leq z \leq l_{0}/2$. The origin lies in the cube center, and the axes are parallel to the cube edges. The iteration rule (\textit{generator}) is to replace the initial cube by eight cubes of edge $\beta_{s}^{(1)}l_{0}$ ($m=1$). The center of the eight cubes are shifted from the origin by the vectors $\bm{a}_{j}=\{\pm \beta_{t}^{(1)}l_{0}, \pm \beta_{t}^{(1)}l_{0}, \pm \beta_{t}^{(1)}l_{0}\}$ with all the combinations of the signs, where $\beta_{t}^{(1)} \equiv (1-\beta_{\mathrm{s}}^{(1)})/2$ and $\beta_{\mathrm{s}}^{(1)}$ is a dimensionless positive parameter for the first iteration, obeying the condition $0 < \beta_{\mathrm{s}}^{(1)} < 1/2$. The second iteration ($m=2$) is obtained by performing an analogous operation to each cube of the first iteration and with the same scaling factor $\beta_{\mathrm{s}}^{(1)}$. For each subsequent iterations we repeat the same operation but for $m=3$ and $m=4$ we take the scaling factor $\beta_{\mathrm{s}}^{(2)}$, for $m=5$ and $m=6$ we take the scaling factor $\beta_{\mathrm{s}}^{(3)}$ and so on. If one consider that the edge of the removed parallelepiped at iteration $m$ is 
\begin{equation}
\gamma_{m}=\alpha^{p_{m}},
\label{eq:removededge}
\end{equation}
where $0 < \alpha < 1$ and the exponent $p_{m}$ is defined as
\begin{equation}
p_{m} \equiv \left \lfloor \frac{1+m}{2} \right \rfloor,
\label{eq:exponent}
\end{equation}
where $m=1,2,\cdots$, and the symbol $\left \lfloor ~ \right \rfloor$ stands for the floor function,
then the scaling factor at the $m$th iteration can be written such as
\begin{equation}
\beta_{\mathrm{s}}^{(m)}=\frac{1-\gamma_{m}}{2}.
\label{eq:scalingfactors}
\end{equation}
The characteristics of the model, together with Eq.~(\ref{eq:removededge}) show that at the $m$th iteration the number of cubes is
\begin{equation}
N_{m}=8^{m}.
\label{eq:number}
\end{equation}
The side length of each cube is given by
\begin{equation}
l_{m}=\frac{l_{0}}{2^{m}}\prod_{i=1}^{m}(1-\gamma_{i}).
\label{eq:sidelength}
\end{equation} 
Therefore, the components of the $\bm{a}_{j}$ vectors, for arbitrarily $m$, can now be written as
\begin{equation}
\beta_{t}^{(m)}=l_{m-1}\frac{1+\gamma_{m}}{4}.
\label{eq:acomps}
\end{equation}

\begin{figure}
\begin{center}
\includegraphics[width=\columnwidth]{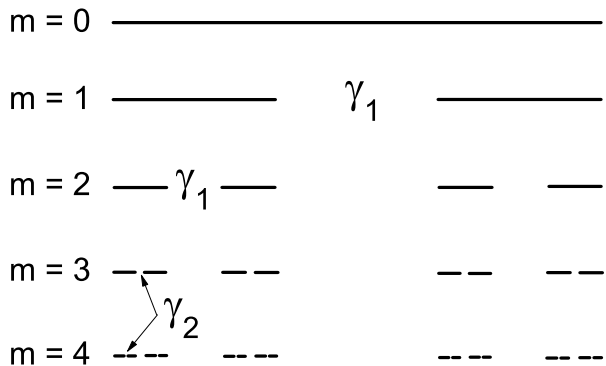}
\end{center}
\caption{
First five iterations for $1D$ fat fractal with $\alpha=1/3$. The dimensionless parameter $\gamma_{m}$ and the scaling factor $\beta_{\mathrm{s}}^{(m)}$ are decreased and, respectively, increased at every second iteration.
}
\label{fig:fig1}
\end{figure}

The fractal dimension of the set can be determined, in the limit of large number of iterations, from relation (using Eqs.~(\ref{eq:number} and~\ref{eq:sidelength}))
\begin{equation}
\mathrm{D}=\lim_{m \to +\infty}{\frac{\ln N_{m}}{\ln (l_{0}/l_{m})}}=3.
\label{eq:fdfat}
\end{equation}
In addition, denoting $v_{1}$ the volume removed at $m=1$, $v_{2}$ the relative volume removed at $m=2$ and so on, we find that the total volume remaining after the $m$th iteration (Lebesgue measure) is $V_{m}=\prod_{i=1}^{m}{(1-v_{i})}$, which has the property that $V_{m} > 0$ if $\sum_{i=1}^{\infty}{v_{i}} < \infty$ and therefore the model fulfills the defining properties of fat fractals~\cite{farmer85,umberger85}. Fig.~\ref{fig:fig1} shows the construction process for the first five iterations of $1D$ projection of the fat fractal at $\alpha=1/3$. 

According to Eq.~(\ref{eq:exponent}), the dimensionless parameter $\gamma_{m}$ and the scaling factor are decreased and, respectively, increased at every second iteration. Since the construction assumes equal values of the scaling factors for two consecutive iterations, the structure is in fact a deterministic thin fractal structure in this ``range" of iterations, each one having a different fractal dimension, given by~\cite{chernyJACR10}
\begin{equation}
D_\mathrm{m}(=\tau)=-\frac{3 \ln 2}{\ln \beta_{s}^{(m)}}.
\label{eq:fd}
\end{equation}

\section{\label{structure}{Fractal form and structure factor}}

We consider as basic subunits of the fractal, cubes with initial edge length $l_{0}$. Therefore, the subunit form factor can be written as~\cite{svergun87:book}
\begin{equation}
F_\mathrm{0}(\bm{q}l_0)=
\frac{\sin(q_{x}l_{0}/2)}{q_{x}l_{0}/2}\frac{\sin(q_{y}l_{0}/2)}{q_{y}l_{0}/2}\frac{\sin(q_{z}l_{0}/2)}
{q_{z}l_{0}/2}.
\label{eq:cubeformfactor}
\end{equation}

In order to calculate the fractal form factor, we apply here the analytical method developed in \cite{chernyJACR10,chernyPRE11} for calculating the fractal form factor of thin fractals. We could use, in principle, the standard Debye formula~\cite{svergun87:book} but its application to deterministic fractals can be cumbersome even for iteration as low as $m=3$, since the number of subunits in the fractal increase exponentially with the iteration number. 

The fractal form factor at the $i$th generation is calculated analytically by means of the generative function, which is determined by the positions of the centers of cubes inside the fractal for each iteration. We consider that the position correspond to a GCF~\cite{chernyJACR10} structure, and therefore the generative function reads as
\begin{equation}
G_{i}(\bm{q})=\cos(q_{x}u_{i})\cos(q_{y}u_{i})\cos(q_{z}u_{i}),
\label{eq:gf}
\end{equation}
where $G_{0}(\bm{q}) \equiv 1$ and the coefficients are given by
\begin{equation}
u_{i}=l_{0}\beta_{\mathrm{t}}^{(i)}\prod_{j=1}^{i-1}\beta_{\mathrm{s}}^{(j)}.
\label{eq:ucoefs}
\end{equation}
The coefficients $u_{i}$ properly takes into account, both, the sizes of subunites through $\beta_{\mathrm{s}}^{(i)}$ and, respectively, their position through $\beta_{\mathrm{t}}^{(i)}$. Then we can write the fractal form factor as
\begin{equation}
F_{m}(\bm{q})=F_{0}(\bm{q}\prod_{i=1}^{m}\beta_{\mathrm{s}}^{(i)})\prod_{i=1}^{m}G_{i}(\bm{q}u_{i}).
\label{eq:finalff}
\end{equation}
Finally, by introducing Eq.~(\ref{eq:finalff}) into Eq.~(\ref{eq:intensitygeneral}), and averaging according to Eq.~(\ref{aver}), the normalized scattering intensity can be written as
\begin{equation}
I_{m}(q)/I_{m}(0)=\langle |F_{m}(\bm{q})|^{2} \rangle.
\label{eq:finalsas}
\end{equation}
The Fourier component of the density of cubes centers are obtained from Eq.~(\ref{eq:fractalff}) and Eq.~(\ref{eq:finalff})
\begin{equation}
\rho_{\bm{q}}=N_{m}\prod_{i=1}^{m}G_{i}(\bm{q}u_{i}).
\label{eq:fc}
\end{equation}
Then, the structure factor is obtained by introducing Eq.~(\ref{eq:fc}) into Eq.~(\ref{eq:sfdef}) which results in
\begin{equation}
S(q)=N_{m} \left \langle  \prod_{i=1}^{m}|G_{i}(\bm{q}u_{i})|^{2} \right \rangle.
\label{eq:sffinal}
\end{equation}

\section{\label{results}{Results and Discussion}}

\begin{figure}
\begin{center}
\includegraphics[width=0.95\columnwidth]{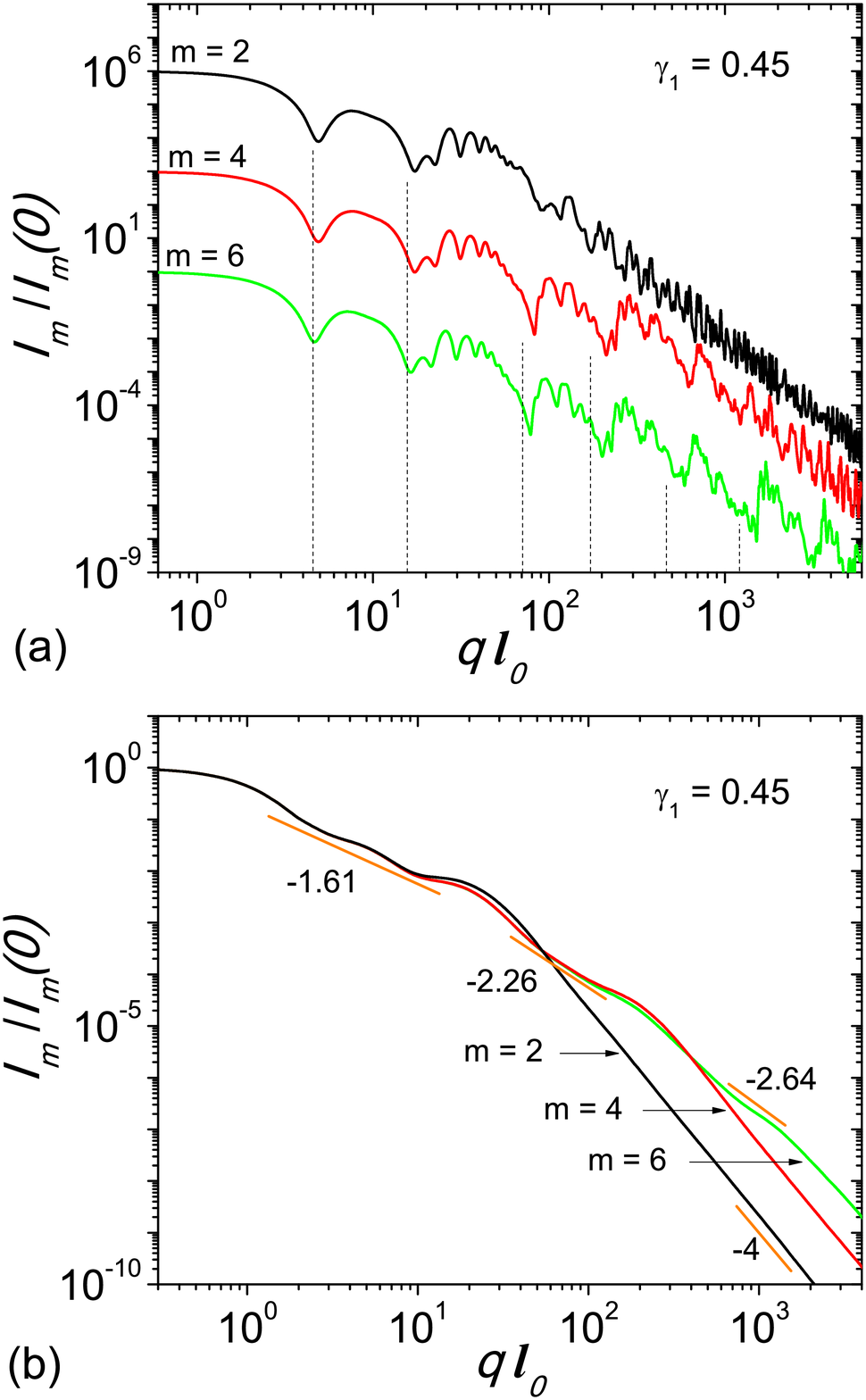}
\end{center}
\caption{
Scattering intensities (Eq.~(\ref{eq:finalsas})) for the second, fourth and sixth iterations of the fat fractal composed of cubes with initial edge length $l_{0}$; a) Intensity from monodisperse fractals (the position of minima indicated by vertical dotted-lines are estimated from Eq.~(\ref{eq:minimaconditions})). The values of the scattering intensity for $m=4$ and $m=2$ are scaled up for clarity with a factor of $10^{3}$ and, respectively $10^{6}$. b) Intensity from polydisperse fractals with $\sigma_{r}=0.4$ (Eq.~(\ref{eq:lengthandvariance})). 
}
\label{fig:fig2}
\end{figure}

\begin{figure}
\begin{center}
\includegraphics[width=0.95\columnwidth]{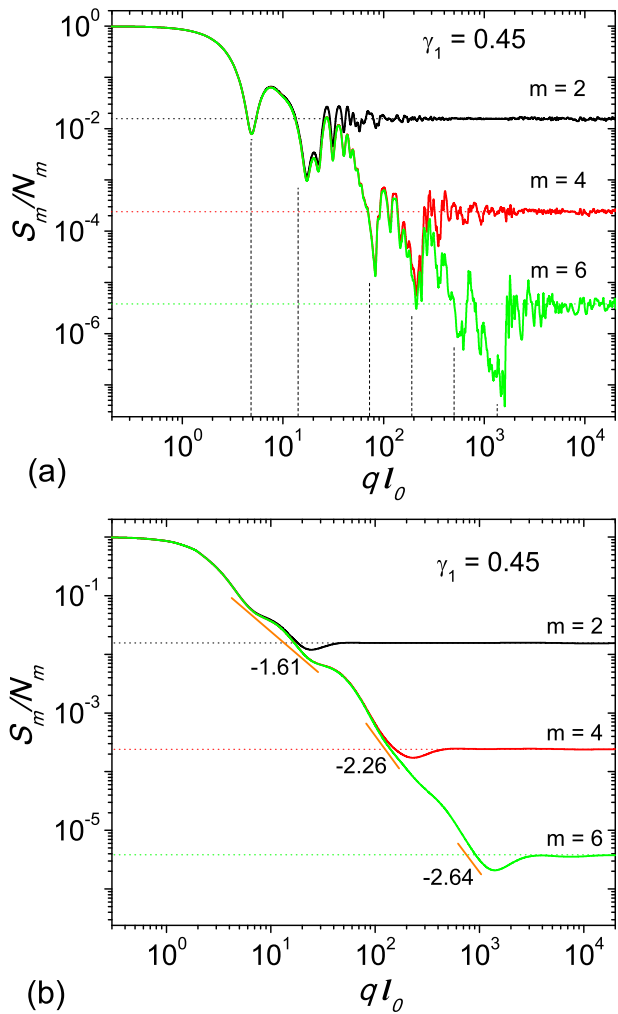}
\end{center}
\caption{
The fractal structure factor (Eq.~(\ref{eq:sffinal})) for the second, fourth and sixth iterations of the fat fractal composed of cubes with initial edge length $l_{0}$, in units of $N_{m}$ (Eq.~(\ref{eq:number})). a) Monodisperse fractals (the position of minima indicated by vertical dotted-lines are estimated from Eq.~(\ref{eq:minimaconditions})); b) Polydisperse fractals with $\sigma_{r}=0.4$ (Eq.~(\ref{eq:lengthandvariance})). The horizontal dotted-lines represent the asymptotes of the structure factor.
}
\label{fig:fig3}
\end{figure}

The numerical results for several iterations for both monodisperse and polydisperse fractal form and structure factor, at fixed $\gamma$, are shown in Fig.~(\ref{fig:fig2}) and, respectively, in Fig.~(\ref{fig:fig3}). At low and intermediate values of $q$ the common feature for the scattering intensities is the appearance of all the regions as seen in SAS experimental data: Guinier at low $q$ and a succession of power-law regimes with decreasing values of the scattering exponents at intermediate $q$. However, at high $q$, the scattering intensity (Eq.~(\ref{eq:finalsas})) shows a Porod region while the fractal structure factor (Eq.~(\ref{eq:sffinal})) shows an asymptotic region.

The Guinier region (a plateau on a double logarithmic scale) is determined by the overall fractal size and can be seen in the region where $ql_{0} \lesssim 1$. In this region the scattering intensity can be approximated by~\cite{svergun87:book}
\begin{equation}
I(q)=I(0)(1-q^{2}R_{\mathrm{g}}^{2}/3+\cdots).
\label{eq:rg}
\end{equation}
The fractal radius of gyration $R_{\mathrm{g}}^{(m)}$ at the $m$th iteration can be determined by expanding the form factor given by Eq.~(\ref{eq:finalff}) in power series in $ql_{0}$ and substituting the result in Eq.~(\ref{eq:finalsas}). Therefore, we obtain
\begin{equation}
R_{\mathrm{g}}^{(m)}=\sqrt{R_{\mathrm{g0}}^{2}\prod_{i=1}^{m}\left (\beta_{\mathrm{s}}^{(i)}\right)^{2}+3l_{0}^{2}\sum_{i=1}^{m}\left(\beta_{\mathrm{t}}^{(i)}\right)^{2}\prod_{k=1}^{i-1}\left(\beta_{\mathrm{s}}^{(k)}\right)^{2}},
\label{eq:rgfinal}
\end{equation}
where $R_{\mathrm{g0}}=l_{0}/2$ for a uniform cube. When all the scaling factors $\beta_{\mathrm{s}}^{(i)}$ and the coefficients $\beta_{\mathrm{t}}^{(i)}$ are equal, Eq.~(\ref{eq:rgfinal}) reduces to the well-known expression for the radius of gyration of thin Cantor fractals~\cite{chernyJACR10}.

The succession of power-law regimes (the fractal region of the fat fractal) is determined by the maximal and minimal distances between the cube centers. Since the smallest distances are of the order of $u_{m}$ (Eq.~(\ref{eq:ucoefs})) then, the beginning of the first power-law regime and the end of the last power-law regime will be found in
\begin{equation}
1 \lesssim ql_{0} \lesssim l_{0}/u_m.
\label{eq:fractalregion}
\end{equation}
In particular, for the monodisperse case (Fig.~(\ref{fig:fig2}a) and Fig.~(\ref{fig:fig3}a)), we have a succession of generalized power-law decays, while for the polydisperse case (Fig.~(\ref{fig:fig2}b) and Fig.~(\ref{fig:fig3}b)), one obtains a succession of simple power-law decays common to experimental SAS data, where the minima and maxima are smeared out~\cite{chernyJACR10,chernyPRE11}. Then the scaling factor at each structural level (iterations with constant scaling factor) can be determined from the periodicity in double logarithmic scale of the quantity $I(q)q^{D_{\mathrm{m}}}$ \textit{vs.} $q$ while the number of fractal iteration can be obtained from the number of periods of the function $I(q)q^{D_{\mathrm{m}}}$~\cite{chernyPRE11}.

In the fractal region of the fat fractal, the structure factor (Eq.~(\ref{eq:sfdef})) approximates very well the scattering intensity (Eq.~(\ref{eq:intsq})), since in this region $F_{0}(\bm{q}\prod_{i=1}^{m}\beta_{\mathrm{s}}^{(i)}) \simeq 1$. The position of minima are obtained when the cubes inside the fractal interfere out of phase, and since the most common distances between the center of mass of the cubes are given by $2u_{m}$, we have the condition $2u_{m}=\pi/q$, which gives the position for the minima (vertical lines in Fig.~(\ref{fig:fig2}a) and Fig.~(\ref{fig:fig3}a)
\begin{equation}
q_{k}l_{0}\simeq \frac{\pi}{2\beta_{\mathrm{t}}^{(k)}\prod_{i=1}^{k}{\beta_{\mathrm{s}}^{(i)}}},
\label{eq:minimaconditions}
\end{equation}
with $k=1,\cdots,m$.

The scattering intensities in Fig.~(\ref{fig:fig2}) and Fig.~(\ref{fig:fig3}) have three main characteristics specific to fat fractal structures, which result from the property that the values of the scaling factors increase with iteration number, according to Eq.~(\ref{eq:scalingfactors}). First, the length of each subsequent power-law regime, in momentum space, decreases. This behavior can be clearly seen in Fig.~(\ref{fig:fig2}b) and Fig.~(\ref{fig:fig3}b). Second, the transition between consecutive power-law regimes is through a ''knee". This is due to the fact that the values of the coefficients $\beta_{\mathrm{t}}$ in Eq.~(\ref{eq:acomps}), which are responsible for the distances between cubes (see Eq.~(\ref{eq:gf})), also decrease. This is in contrast with scattering from multiphase systems, where the ''knee" position depends on the scattering length density of each component phase~\cite{cherny12,chernyArxiv}. Third, since the values of $\beta_{\mathrm{s}}^{(m)}$ depend on the initial value $\beta_{s}^{(1)}$, the fractal dimension for each range can be determined only by specifying the fractal dimension at $m=1,2$ and the value of $\gamma_{1}$ in Eq.~(\ref{eq:removededge}).  

Beyond the last power-law regime (or the last GPLD in the monodisperse case) we have $q \gtrsim 1/u_{m}$. In this region the fractal structure factor in Eq.~(\ref{eq:sffinal}) is $S(q)\simeq 1$~\cite{chernyPRE11} and therefore we find that the asymptotic values tend to $1/N_{m}$ (Fig.~(\ref{fig:fig3}a) and Fig.~\ref{fig:fig3}b)), as for the case of thin fractals~\cite{chernyJACR10,chernyPRE11}. From another hand, in this region, the scattering intensity (Eq.~(\ref{eq:finalsas})) follows the Porod law (Fig.~(\ref{fig:fig2}a) and Fig.~\ref{fig:fig3}b)), since the size of the initiator is of the same order as $l_{0}$. The beginning of the Porod region allows us to obtain the size of the smallest unit (here cube) constituting the fractal.

\section{\label{conclusions}{Conclusions}}

We develop a fat fractal model based on an extension of the generalized Cantor fractal~\cite{chernyJACR10}, and which allows us to explain experimental SAS data showing a succession of power-law regimes (or GPLD regimes in the monodisperse case) with arbitrarily decreasing values of the scattering exponents in the range from zero to three. The main feature of the model, which allows to explain such type of data, is the increase of the scaling factor after a given number of iterations (here, every second iteration) which, in turn, implies arbitrarily decreasing values of the fractal dimensions. We derive an analytical expression for the form and structure factor, which describe scattering from non-interacting, mono- and polydisperse, randomly oriented, $3D$ fat fractals. We have calculated analytically the radius of gyration of the fractal.

We have shown that the present analysis allows us to obtain three main structural characteristics of fat fractals. First, the edges of all the fractal regions (through the positions of minima in Eq.~(\ref{eq:minimaconditions})), which can independently be controlled by choosing various expressions for the floor function $p_{m}$ defined in Eq.~(\ref{eq:exponent}). Second, the fractal dimensions and the scaling factors corresponding to \textit{each} structural level, which can be controlled by the parameter $\alpha$ in Eq.~(\ref{eq:removededge}). Third, the number of particles composing the fractal, from the asymptote of the structure factor in Eq.~(\ref{eq:sffinal}), and which can be controlled by a different definition of the iterative operation (generator). In addition, from the calculated radius of gyration (Eq.~(\ref{eq:rgfinal})) and from the scattering intensity (Eq.~(\ref{eq:finalsas})) one can obtain information about the overall size of the fractal and respectively, about the sizes of the smallest units composing the fractal.

The model could serve to describe and analyze growth phenomena of biological objects or clusters at nano- and micro scales.

\begin{acknowledgments}
The author is grateful to Alexander Yu. Cherny, Vladimir A. Osipov and Alexander I. Kuklin for valuable discussions. Financial support from JINR–--IFIN-HH projects is acknowledged.
\end{acknowledgments}

\bibliography{sasfatf}

\end{document}